\documentclass[%
 reprint,
nofootinbib,
aps,
prd,
]{revtex4-2}

\usepackage{graphicx}
\usepackage{dcolumn}
\usepackage{bm}

\usepackage{graphicx}
\usepackage{tikz}
\usetikzlibrary{matrix}

\usepackage{setspace}
\usepackage{xcolor}
\usepackage{hyperref}
\usepackage{mathtools}
\usepackage{esint}
\usepackage{amssymb}
\usepackage{eucal}
\usepackage{graphicx}
\usepackage{amsmath}
\usepackage{mathtools}
\usepackage{braket}
\usepackage{amsmath}
\usepackage{bbold}

\begin{document}


\title{\texorpdfstring{$T\bar T$}{} deformation of Calogero-Sutherland model \texorpdfstring{\\}{} via dimensional reduction}

\author{Dmitriy Pavshinkin}%
 \email{dmitriy.pavshinkin@phystech.edu}
\affiliation{%
 Moscow Institute of Physics and Technology, Dolgoprudny 141700, Russia}%


\begin{abstract}
We perform the dynamical change of coordinates to derive a generalization of the trace relation and apply it to the non-linear Schr\"odinger model. After that, we work out the dimensional reduction of the bilinear $T\bar T$ operator and obtain the new $T\bar T$-like deformation of the quantum mechanics of free non-relativistic fermions and interacting Calogero-Sutherland particles. The deformation modifies the spectrum of the Hamiltonian but does not alter its eigenfunctions. The deformed classical Lagrangian is also obtained. Finally, we study a particular deformation of the two-dimensional Yang-Mills theory that maps the gauge theory onto a system of $T\bar T$-perturbed fermions.

\end{abstract}

\maketitle

\section{Introduction}\label{sec:level1}

The significant progress in studying $T\bar T$ relativistic quantum field theories  \cite{SmirZam,Tateo,Jiangreview} has motivated the recent formulation of $T\bar T$-like deformations for the more general case of non-Lorentz invariant QFTs \cite{Cardy} as well as integrable spin chains \cite{spin3,spin4,spin1,spin2} and many-body systems \cite{CardyDoyon,jiang,nonrelativ}. The latter has aided in identifying prominent landmarks of exactly solvable irrelevant deformations. The authors of \cite{CardyDoyon,jiang} have shown that deformed non-relativistic many-body theories also become non-local and exhibit the Hagedorn behavior in a density of states, sharing these properties with string theory, yet they are still solvable. The qualitative picture underlying these phenomena has been perceived as stretching point particles to a finite width, determined by a system's total energy.

In \cite{jiang}, the $T\bar T$ deformation of the integrable Lieb-Liniger (LL) model, describing 1d  Bose particles with repulsive delta interaction, was explored.  The deformation was defined on the level of the Bethe ansatz equations as well as the flow of the Hamiltonian.
In the finite volume, $T\bar T$ was involved in a deformation of both the Hamiltonian's spectrum and eigenfunctions.
On the other hand, the S-matrix in the infinite volume was accompanied by the CDD factor, resulting in an appropriate deformation of the Bethe equations.
At the same time, the authors of \cite{NLS} studied the deformation of the non-linear Schrödinger (NLS) model -- the classical-field-theory prototype of LL. As a result,  they found a solution employing the dynamical (field-dependent) change of coordinates method, initially formulated for the relativistic field theories \cite{Change}. The method suggests that one can obtain a non-perturbative solution to the Lagrangian flow equation as a first-order perturbed original Lagrangian, written in a particular coordinate system.

Besides the one that uses a bilinear operator \cite{CardyDoyon,jiang}, there is an equally natural definition of the $T\bar T$-perturbed quantum mechanics \cite{gross1,gross2} that is inspired by the holographic interpretation of $T\bar T$-CFT$_2$ \cite{Moving}. This deformation is obtained by the dimensional reduction of the 2d $T\bar T$ operator, provided the $T\bar T$-CFT$_2$ trace relation is imposed. Therefore, the Schwarzian QM, deformed in this way, becomes dual to finite radial cutoff AdS$_2$.
A sharp difference between the deformation by the bilinear $T\bar T$ operator and the dimensionally reduced one is that the deformed Hamiltonian is a function of the original Hamiltonian in the latter case.  As a result, eigenfunctions do not change, and  correlation functions are fixed by the deformed spectrum.

This work aims to study $T\bar T$-like deformations of classical and quantum non-relativistic Calogero-Sutherland-Moser model \cite{Calogero,Sutherland,Moser}. The CSM is at the heart of integrable systems and has been unearthed all over theoretical physics: in CFT \cite{vasiliev,Cardy1}, condensed matter \cite{solitons}, and random matrix theory \cite{rmt}, to name just a few. In the following, we are mainly focusing on its relation to gauge theories. The CSM describes a system of identical particles living on a circle and interacting in a specific way. A distinguishing property of 1d interacting many-body systems is an indissoluble relation between their pairwise statistical and dynamical interactions. In this regard, the CSM demonstrates fractional exclusion statistics and turns out to be related to two-dimensional anyons \cite{Anyons}. The formulation of the CSM in the non-relativistic field theory formalism, however, remains unclear. 

Utilizing the field-dependent change of coordinates method, we derive the trace and determinant flow of a stress-energy tensor and obtain a generalized trace-det relation. Next, we apply this relation to the on-shell configuration of the NLS model in the non-interacting limit. It helps us find a dimensionally reduced $T\bar T$ operator corresponding to the QM of both free bosons and fermions.  The presented strategy is similar to obtaining the $T\bar T$ deformation of conformal QM in \cite{gross1}. Accordingly, the new deformation can be considered as its ``non-relativistic'' counterpart. Then, we determine the deformation of CSM. By construction, the obtained operator does not change the eigenfunctions and, in general, is not equivalent to the bilinear deformation. 

Another exciting feature of integrable many-body systems is their relation to gauge theories, a duality between a phase space of particles and topological gauge degrees of freedom (see \cite{gaugereview} for review).  The two-dimensional Yang-Mills is the simplest example of this connection. We consider YM on a cylinder with $L$-periodic spatial coordinate and with the $U(N)$ gauge group. Because of the almost topological nature of 2d YM, its Hilbert space consists only of a finite number of degrees of freedom. These effective DOFs correspond to eigenvalues of the spatial Wilson loop that resemble a system of $N$ free non-relativistic fermions on a circle of length $L^{\prime}\sim L^{-1/2}$ \cite{MinPol,douglas}. Furthermore, if the Wilson line in a particular representation is added along the cylinder, the underlying phase space dynamics becomes that of CSM particles \cite{gn1,gn2,CalWL}.

The $T\bar T$ deformation of the YM theory has been studied in various contexts \cite{BornInf,tierz1,DBI,Shyam,add1} by now. Also, new phenomena have been discovered \cite{GPT,tierz2}. The clear particle-like interpretation, though, seems to have been lost after the deformation. The on-shell $T\bar T$ operator consistent with the YM is reduced to $(T^t_{\,t})^2$, and the resulting theory completely differs from a $T\bar T$-perturbed system of non-relativistic particles.
That is a consequence of the nontrivial mapping $L^{\prime}\sim L^{-1/2}$ described above. However, the new solvable deformation we propose in the paper maps the 2d YM to a system of generalized hard-rod fermions.

The rest of the article is structured as follows.
The next section goes over the main formulas concerning the CSM model that we use during the paper. Then we review the $T\bar{T}$ bilinear deformation of integrable many-body systems and obtain the deformed CSM energy spectrum. Section 3 is devoted to the derivation of dimensionaly reduced $T\bar{T}$ operator corresponding to Calogero particles.
 In section 4 we obtain a certain deformation of classical and quantum Yang-Mills theory that maps the gauge theory onto a system of $T\bar{T}$-perturbed fermions. The Hamiltonian reduction of the deformed YM on a cylinder with an additional Wilson line is derived in section 5.
In Discussion section, we propose the most interesting extensions of our study.

\section{Preliminaries}
\subsection{Calogero-Sutherland model}
The classical CSM Hamiltonian is defined as follows
\begin{equation}
H^{\text{\tiny{CSM}}}=\frac{1}{2m}\sum_{i=1}^Np_i^2+\frac{1}{2m}\Big(\frac{\pi}{{L}}\Big)^2\sum_{i<j}^N\frac{\gamma^2}{\sin^2\frac{\pi}{{L}}(q_i-q_j)}
\end{equation}
where $\gamma$ is the coupling constant, and $\{p_i,q_i\}$ are canonically conjugate variables describing momenta and position of $N$ non-relativistic particles on a circle of length $L$ (see \cite{CalLect,PolyLect} for review).
The classical integrability of the system implies the existence of $N$ commuting integrals of motion $H_k$, which can be reproduced in a unified fashion from the Lax matrix.

The Hamiltonian is quantized by replacing the classical momentum  $p_k\rightarrow -i\hbar\frac{\partial}{\partial q_k}$ and performing the additional quantum shift $\gamma^2\rightarrow\gamma(\gamma-\hbar)$:
\begin{equation}
\hat{H}^{\text{\tiny{CSM}}}=-\frac{\hbar^2}{2m}\Big[\sum_{i=1}^N\frac{\partial^2}{\partial q_i^2}-\Big(\frac{\pi}{{L}}\Big)^2\sum_{i<j}^N\frac{\beta(\beta-1)}{\sin^2\frac{\pi}{{L}}(q_i-q_j)}\Big],
\end{equation}
Here we have defined the quantum coupling constant $\beta=\frac{\gamma}{\hbar}$.
From now on, we set $\hbar^2=m=1$. 
 The ground state wave function is given by
\begin{equation}
\tilde{\Delta}^{\beta}({q})=\prod_{i<j}^N\Big(\frac{L}{\pi}\sin\frac{\pi}{L}(q_i-q_j)\Big)^{\beta}.
\end{equation}
It also defines a particle position probability distribution that is consistent with the joint probability densities for the eigenvalues of Dyson's circular ensembles: $2\beta= 1, 2, 4$.
The wavefunction appropriate to describe excited states is
\begin{equation}
\Psi^{\text{\tiny{CSM}}}=J^{\beta}_{\boldsymbol{n}}(q)\tilde{\Delta}^{\beta}(q).
\end{equation}
Here the $J^{\beta}_{\boldsymbol{n}}$ denote the Jack symmetric functions. They are indexed by partitions $\boldsymbol{n}$ and reduce to the Schur polynomials at $\beta=1$. The eigenvalue associated with the set of integers $\{n_1\geq n_2...\geq n_N\}$ is given by
\begin{multline}
E(\boldsymbol{n})= \frac{1}{2}\Big(\frac{\pi}{L}\Big)^2\sum_{j=1}^Nn_j\big(n_j+\beta(N+1-2j)\big)\\
 +\beta^2\Big(\frac{\pi}{L}\Big)^2\frac{N(N^2-1)}{24}
\end{multline}
while the ground state energy matches to $\boldsymbol{n}=0$.

The manifestation of CSM's integrability lies in a quasi-particle interpretation of the energy spectrum.
Introducing the quasi-particle momenta
 \begin{equation}
\tilde{p}_j=\frac{2\pi}{L}\big(n_{\scriptstyle N+1-j}\displaystyle+\beta(j-\scriptstyle\frac{N+1}{2}\displaystyle)\big)
\end{equation}
we obtain
 \begin{equation}
E(\boldsymbol{n})=\frac{1}{2}\sum_{j=1}^N\tilde{p}_j^2,~~P(\boldsymbol{n})=\sum_{j=1}^N\tilde{p}_j.
\end{equation}
These quasi-momenta exhibit the generalized selection rule
 \begin{equation}
\tilde{p}_i-\tilde{p}_j\geq 2\pi\beta/L,
\end{equation}
reproducing the bosonic statistics at $\beta=0$  and fermionic one at $\beta=1$.
Therefore, the CSM 
features both anyonic statistics and dynamical interaction.

The spectrum of the trigonometric CSM can be obtained from the Bethe Ansatz equations for the rational model.
The rational CSM has the constant phase shift
\begin{equation}
\theta(p_i-p_j)=\pi(\beta-1)\text{sign}(p_i-p_j)
\end{equation}
where $p_i$ are asymptotic momenta.
 Provided $L$-periodic boundary conditions are imposed on wave functions, the BA equations take the following form  
 \begin{equation}
p_j=\frac{2\pi I_j}{L}+\frac{\pi(\beta-1)}{L}\sum_{k\neq j}^N\text{sign}(p_j-p_k).
\end{equation}
 Assuming the ordering $p_1<...<p_N$ one obtains
  \begin{equation}\label{2.11}
p_j=\frac{2\pi I_j}{L}+\frac{\pi(\beta-1)}{L}(2j-N-1).
\end{equation}
The $p_j$ in the formula above coincide with the quasi-momenta in the trigonometric CSM after the identification
\begin{equation}
n_{\scriptstyle N+1-j}\displaystyle =I_j+\frac{N+1}{2}-j.
\end{equation}

\subsection{\texorpdfstring{$T\bar T$}{}-perturbed non-relativistic integrable systems}

The deformation of non-relativistic integrable models can be studied in a universal fashion by means of the Betha Ansatz technique.
In the infinite volume limit, the deformation
modifies the S-matrix by the CDD-like phase factor
\begin{equation}
S_{\lambda}(u,v)=e^{-i\lambda[p(u)e(v)-p(v)e(u)]}S(u,v)
\end{equation}
where $u$, $v$ are rapidities.
Therefore, under the condition of the non-relativistic dispersion relations, the deformed BA equations in the finite volume take the form
 \begin{equation}
p_j[L-\lambda E_{\lambda}(p)]+\lambda p_j^2 P_N+\sum^N_{k\neq j}\theta(p_j,p_k)=2\pi I_j
\end{equation}
Obtaining the deformed spectrum in the sector of zero total momentum is fairly straightforward. The deformation changes the size of the system $L\rightarrow L-\lambda E_{\lambda}$. 
In particular, using \eqref{2.11} with $I_j=j-\frac{N+1}{2}$, we find the ground state energy of the $T\bar{T}$-CSM
\begin{equation}
E_{\lambda}=\frac{1}{2}\sum_{j=1}^Np_j^2=\frac{\beta^2\pi^2}{(L-\lambda E_{\lambda})^2}\frac{N(N^2-1)}{24}.
\end{equation}
An interesting qualitative picture behind the deformation is that a system of particles in a reduced volume can be interpreted as a system of hard rods, each of which has a length equal to the excluded volume divided by the number of particles. 

The $T\bar{T}$ deformation can also be formulated without relying on integrability conditions using a one-parameter family of Hamiltonians satisfying the equation
\begin{equation}
\frac{d H_{\lambda}}{d\lambda}=\int dxT\bar{T}_{\lambda}(x,t).
\end{equation}
At the classical level, this definition is equivalent to the Lagrangian flow equation (see \eqref{3.1}).
The authors of \cite{spin1,KP} proposed the following rewriting of the spatial integral over the $T\bar{T}$ operator:
\begin{equation}\label{17}
\int dxT\bar{T}_{\lambda}(x,t)=i[H_{\lambda}(t),\mathcal{X}_{\lambda}(t)]+\mathcal{Y}_{\lambda}(t)
\end{equation}
The first term is defined by the bilocal operator 
\begin{equation}
\mathcal{X}_{\lambda}(t)=\int_{x<y}dxdy \,T^t_{\,t}(x,t)T^t_{\,x}(y,t)
\end{equation}
The commutator is responsible for the canonical transformation of the Hamiltonian and its eigenfunctions
\begin{equation}
H\rightarrow U_{\lambda}HU_{\lambda}^{-1},~~
|\Psi\big>\rightarrow U_{\lambda}|\Psi\big>.
\end{equation}
Notice that, considering a theory of free particles with 
$
H^{\text{free}}=\frac{1}{2}\sum_{i=1}^Np_i^2
$,
one can easily make sure that the commutator does not affect the Hamiltonian 
$
[H_{\lambda}^{\text{free}},\mathcal{X}_{\lambda}
]=0.
$

The second term in \eqref{17} is
\begin{multline}\label{2.20}
\mathcal{Y}_{\lambda}(t)= \big({\boldsymbol{P}_{\lambda}}^t_{\,x}{\boldsymbol{P}_{\lambda}}^x_{\,t}-{\boldsymbol{P}_{\lambda}}^t_{\,t}{\boldsymbol{P}_{\lambda}}^x_{\,x}\big)/L,\\{\boldsymbol{P}_{\lambda}}^{\alpha}_{\,\beta}=\int dx {{T}_{\lambda}}^{\alpha}_{\,\beta}(x,t).
\end{multline}
It results in the flow of the energy spectrum
\begin{equation}
\frac{d E_{\lambda}}{d\lambda}=\langle\mathcal{Y}_{\lambda}\rangle.
\end{equation}
In the infinite volume $L\rightarrow\infty$, the operator $\mathcal{Y}_{\lambda}$ is suppressed, and the unitary matrix 
\begin{equation}
U_{\lambda}=P\exp\big\{-i\int_0^{\lambda^{}}d\lambda^{\prime}\mathcal{X}_{\lambda^{\prime}}\big\}.
\end{equation}
 
We introduce the following notation for the bilinear $T\bar{T}$ deformation of 1d theories
\begin{equation}\label{2.22}
\mathcal{O}^{T\bar{T}}_{\lambda}\coloneqq\int dxT\bar{T}_{\lambda}(x,t).
\end{equation}

\section{\texorpdfstring{$T\bar T$}{}-QM via dimensional reduction}

In this section, we elaborate the ${T\bar{T}}$-like deformation of QM, the definition of which does not depend on an underlying 2d field theory.
On the example of free bosons and fermions, we derive a composite operator that is a function of the one-dimensional stress-energy tensor. We use the on-shell trace relation to express an auxiliary $T^x_{\,x}$ component of the 2d stress-energy tensor through the others, which have direct interpretations in a 1d theory. Then, using the quasi-particle description, we generalize this deformation to the CSM.
The price of this dimensional reduction is that the resulting deformation only affects the Hamiltonian spectrum and does nothing with its eigenvalues. In this sense, the reduced operator is equivalent to $\mathcal{Y}_{\lambda}$ \eqref{2.20}.

\subsection{Generalized trace relation}

The $T\bar{T}$ deformation of classical 2d field theories was originally defined as the flow equation
\begin{equation}\label{3.1}
\frac{dS_{\lambda}}{d\lambda}=-\int d^2xT\bar{T}_{\lambda}(x,t)
\end{equation}
with the action 
\begin{equation}
S_{\lambda}=\int d^2x{\mathcal{L}}_{\lambda}(x,t).
\end{equation}
It turns out that the deformed Lagrangian can be expressed in terms of the original one as follows
\begin{equation}
\int d^2x{\mathcal{L}}_{\lambda}(\mathbf{x})=\int d^2x^{\prime}\big({\mathcal{L}}_{0}(\mathbf{x^{\prime}})-\lambda \det[T_0(\mathbf{x^{\prime}})]\big)
\end{equation}
where the Jacobian matrix between the coordinate bases $\mathbf{x}^{\alpha}=(x,t)$ and ${\mathbf{x}^{\prime}}^{\alpha}=(x^{\prime},t^{\prime})$ is
\begin{equation}
J^{-1}(\mathbf{x^{\prime}})=\begin{pmatrix}
    1+\lambda {T_0}^x_{\,x}(\mathbf{x^{\prime}})      & -\lambda {T_0}^t_{\,x}(\mathbf{x^{\prime}})
    \\
   -\lambda {T_0}^x_{\,t}(\mathbf{x^{\prime}})&       1+\lambda {T_0}^t_{\,t}(\mathbf{x^{\prime}})
\end{pmatrix}
\end{equation}
Therefore, the deformed Lagrangian density expressed in $\mathbf{x}$ coordinates is
\begin{equation}\label{3.5}
{\mathcal{L}}_{\lambda}(\mathbf{x})=\frac{\mathcal{L}_{0}(\mathbf{x^{\prime}}(\mathbf{x}))-\lambda \det[T_0(\mathbf{x^{\prime}}(\mathbf{x}))]}{\det[J^{-1}(\mathbf{x^{\prime}}(\mathbf{x}))]}.
\end{equation}

Performing the standard formula for a stress-energy tensor, we obtain 
\begin{equation}\label{3.6}
{T_{\lambda}}_{\,\nu}^{\mu}(\mathbf{x})=\frac{\partial\mathbf{x}^{\mu}}{\partial \mathbf{{x}^{\prime}}^{\alpha}}\frac{\partial\mathbf{{x}^{\prime}}^{\beta}}{\partial \mathbf{x}^{\nu}}\frac{{T_{0}}_{\,\beta}^{\alpha}(\mathbf{x^{\prime}}(\mathbf{x}))+\delta_{\,\beta}^{\alpha}\lambda\det T_0(\mathbf{x^{\prime}}(\mathbf{x}))}{\det J^{-1}(\mathbf{x^{\prime}}(\mathbf{x}))}.
\end{equation}
In particular, we find the expression for the deformed Hamiltonian density $\mathcal{H}_{\lambda}\coloneqq{T_{\lambda}}_{\,t}^{t}$ 
\begin{equation}
\mathcal{H}_{\lambda}(\mathbf{x})=\frac{\mathcal{H}_{0}{(\mathbf{x^{\prime}}(\mathbf{x}))+\lambda\det T_0(\mathbf{x^{\prime}}(\mathbf{x}))}}{\det J^{-1}(\mathbf{x^{\prime}}(\mathbf{x}))}.
\end{equation}
Let us underline that the preceding equation is not a closed formula for the Hamiltonian's density (as well as \eqref{3.5}). On the contrary, it holds at the level of the equations of motion, and, after the theory is specified, the $\mathbf{x}^{\prime}$-derivatives of fields included in the $\mathcal{H}_{0}$ must be expressed in the old coordinates $\mathbf{x}^{\alpha}$.

From \eqref{3.6} we obtain the $T\bar{T}$-flow of the trT and detT:
\begin{equation}
{\text{tr}\,T_{\lambda}(\mathbf{x})}=\frac{{\text{tr}\,T_{0}(\mathbf{x^{\prime}}(\mathbf{x}))}+2\lambda\det T_0(\mathbf{x^{\prime}}(\mathbf{x}))}{\det J^{-1}(\mathbf{x^{\prime}}(\mathbf{x}))},
\end{equation}
\begin{equation}
{\det T_{\lambda}(\mathbf{x})}=\frac{{\det T_{0}(\mathbf{x^{\prime}}(\mathbf{x}))}}{\det J^{-1}(\mathbf{x^{\prime}}(\mathbf{x}))}.
\end{equation}

Using these expressions  we come to the generalized trace relation
\begin{equation}\label{3.10}
\text{tr}\,T_{\lambda}=2\lambda\det T_{\lambda}+\frac{{\text{tr}\,T_{0}}}{1+\lambda \text{tr}\, T_0+\lambda^2\det T_0}
\end{equation}
Notice that this formula holds up to total derivatives and should be regarded under the integral. Using the tracelessness condition of a CFT we recover 
\begin{equation}
\text{tr}\,T_{\lambda}=2\lambda\det T_{\lambda},
\end{equation}
which has been obtained in many papers based on the dimensional analysis.

In the next section we apply the trace relation to a specific classical non-relativistic field theory.

\subsection{NLS trace relation}
 Recently, the idea of dynamical change of coordinates has been applied to non-relativistic models on the example of the non-linear Schrödinger (NLS) model with generic potential $V$ \cite{NLS}. Usually, the potential is defined as $V(x_i-x_j)=c\delta(x_i-x_j)$.
Canonical quantization of this model leads to a QFT describing Bose gas with a pairwise delta interaction. In the $N$-particle sector, this QFT is equivalent to the Lieb–Liniger model. 

We are interested in the limit of non-interacting bosons $c\rightarrow 0$. The corresponding Lagrangian density 
\begin{equation}
\mathcal{L}_0=\frac{i}{2}(\psi^*\partial_t\psi-\partial_t\psi^*\psi)-\partial_x\psi^*\partial_x\psi
\end{equation}
and components of the stress-energy tensor
\begin{multline}
{T_0}_{\,t}^t=\partial_x\psi^*\partial_x\psi,~~~
{T_0}_{\,x}^t=\frac{i}{2}(\psi^*\partial_x\psi-\partial_x\psi^*\psi)\\
{T_0}^x_{\,t}=-(\partial_x\psi^*\partial_t\psi+\partial_t\psi^*\partial_x\psi),\\
{T_0}^x_{\,x}=-\frac{i}{2}(\psi^*\partial_t\psi-\partial_t\psi^*\psi)-\partial_x\psi^*\partial_x\psi.
\end{multline}
Equations of motion:
\begin{equation}
i\partial_t\psi^*=\partial^2_x\psi^*,~~-i\partial_t\psi=\partial^2_x\psi
\end{equation}
The crucial observation is that on EoM 
\begin{multline}
{T_0}^x_{\,x}=\frac{1}{2}(\psi^*\partial^2_x\psi+\partial^2_x\psi^*\psi)-\partial_x\psi^*\partial_x\psi\\
=-2\partial_x\psi^*\partial_x\psi+\textstyle\frac{1}{2}\displaystyle\partial_x^2|\psi|^2,
\end{multline}
and the diagonal components of the stress-energy tensor obey the relation
\begin{equation}
{T_0}^x_{\,x}=-2{T_0}^t_{\,t}+\textstyle\frac{1}{2}\displaystyle\partial_x^2|\psi|^2.
\end{equation}

Using this formula along with \eqref{3.6} and \eqref{3.10}, we obtain the on-shell trace relation of non-relativistic bosons:

\begin{equation}\label{40}
\text{tr}\,{T}_{\lambda}=3\lambda\det {T}_{\lambda}-{{T}_{\lambda}}^t_{\,t}+ \frac{\textstyle\frac{1}{2}\displaystyle\partial_x^2|\psi|^2}{1+\lambda \text{tr}\, T_0+\lambda^2\det T_0} .
\end{equation}

Now we are ready to perform the dimensional reduction.

\subsection{\texorpdfstring{${{\mathcal{O}}}^{\text{\tiny{GHR}}}_{\lambda}$}{} deformation}
Consider $T\bar T$ deformation of a field theory living on a cylinder with $L$-periodic space coordinate. The Hamiltonian flow equation is
\begin{equation}
\frac{dH_{\scriptstyle\lambda\displaystyle}}{d\lambda}=\mathcal{O}^{T\bar{T}}_{\lambda}=-\int _0^Ldx \big({T_{\lambda}}^t_{\,t}{T_{\lambda}}^x_{\,x}-{T_{\lambda}}^t_{\,x}{T_{\lambda}}^x_{\,t}\big).
\end{equation}
The recipe behind the dimensional reduction is to eliminate the $x$-dependence by substituting \begin{equation}
{T_{\lambda}}^{\mu}_{\,\nu}(x,t)\rightarrow {\boldsymbol{P}_{\lambda}}^{\mu}_{\,\nu}(t)/L.
\end{equation} 
In the dimensionally reduced theory we have
\begin{equation}\label{3.19}
\frac{dH_{\scriptstyle\lambda\displaystyle}}{d\lambda}=-\int _0^Ldx \big({\boldsymbol{P}_{\lambda}}^t_{\,t}{\boldsymbol{P}_{\lambda}}^x_{\,x}-{\boldsymbol{P}_{\lambda}}^t_{\,x}{\boldsymbol{P}_{\lambda}}^x_{\,t}\big)/L^2.
\end{equation}
On can see that the r.h.s. is nothing but the $\mathcal{Y}_{\lambda}$ operator \eqref{2.20}.
However, in a 1d theory, ${T_{\lambda}}^x_{\,x}$ should be considered as an auxiliary field.\footnote{ In the $T\bar{T}$-perturbed CFT$_1$ considered in \cite{gross1}, ${T_{\lambda}}^x_{\,x}$ was defined as a field dual to the dilaton in the bulk.}
In the case of non-interacting bosons we can use \eqref{40} to find the ${\boldsymbol{P}_{\lambda}}^x_{\,x}$ component:
\begin{equation}
{\boldsymbol{P}_{\lambda}}^x_{\,x}=-\frac{2L{\boldsymbol{P}_{\lambda}}^t_{\,t}+3{\boldsymbol{P}_{\lambda}}^x_{\,t}{\boldsymbol{P}_{\lambda}}^t_{\,x}}{L-3\lambda {\boldsymbol{P}_{\lambda}}^t_{\,t}}+\text{tot.\,der.}
\end{equation}

Substituting this expression in \eqref{3.19}  we obtain the dimensionally reduced operator 
\begin{equation}\label{3.21}
\mathcal{O}_{\lambda}^{\text{\tiny{GHR}}}(t)\coloneqq\frac{2({\boldsymbol{P}_{\lambda}}^t_{\,t})^2+{\boldsymbol{P}_{\lambda}}^x_{\,t}{\boldsymbol{P}_{\lambda}}^t_{\,x}}{L-3\lambda {\boldsymbol{P}_{\lambda}}^t_{\,t}}.
\end{equation}

In quantum mechanics, free non-relativistic bosonic and fermionic particles can be described using the same Hamiltonian, which involves only the kinetic term $\frac{{\hat{p}}^2}{2}$. So the formula above is also applicable to free fermions.
After that, we turn on the trigonometric interaction between particles. Motivated by the quasi-particle interpretation of the spectrum of the CSM, we introduce the ``quasi-quantities'' ${\boldsymbol{{\tilde{P}}}_{\lambda}}^{\mu}_{\,\nu}$ such that
\begin{equation}
{{\boldsymbol{\tilde{P}}}_0}^{t}_{\,t}\coloneqq \frac{1}{2}\sum_{j=1}^N\tilde{p}^2_j=E^{\text{\tiny{CSM}}},~~{{\boldsymbol{\tilde{P}}}_0}^{t}_{\,x}\coloneqq \sum_{j=1}^N\tilde{p}_j
\end{equation}
and define the deformation in the similar way
\begin{equation}
\frac{dH_{\lambda}^{\text{\tiny{CSM}}}}{d\lambda}=\frac{2({\boldsymbol{{\tilde{P}}}_{\lambda}}^t_{\,t})^2+{\boldsymbol{{\tilde{P}}}_{\lambda}}^x_{\,t}{\boldsymbol{{\tilde{P}}}_{\lambda}}^t_{\,x}}{L-3\lambda {\boldsymbol{{\tilde{P}}}_{\lambda}}^t_{\,t}}.
\end{equation}
From now on, for simplicity, we consider the zero momentum sector ${\boldsymbol{\tilde{P}}}^{t}_{\,\,x}=0$.
Notice that the potential in $H_{\lambda}^{\text{\tiny{CSM}}}$ does not commute with $\mathcal{X}_{\lambda}$, and therefore $\mathcal{O}^{\text{\tiny{GHR}}}\neq\mathcal{O}^{T\bar{T}}$. The key difference is that the $\mathcal{O}^{\text{\tiny{GHR}}}$-perturbed Hamiltonian is a function of the original one, and the wave functions are unchanged\footnote{New eigenstates do not appear upon the deformation in the case of a finite number of degrees of freedom.}
\begin{equation}
H_{\lambda}^{\text{\tiny{CSM}}}=f_{\lambda}\big(H_{0}^{\text{\tiny{CSM}}}\big),~~\ket{E_n}_{\lambda}=\ket{E_n}_{0}.
\end{equation}

We corroborate the results by the explicit calculation of the deformed spectrum. Using the Hellmann–Feynman theorem 
we obtain the spectrum flow
\begin{equation}
\frac{dE_{\lambda}}{d\lambda}=\frac{2E_{\lambda}^2}{L-3\lambda E_{\lambda}}.
\end{equation}
Multiplying this equality by $(L-3\lambda E_{\lambda})(L-\lambda E_{\lambda})$ we get 
\begin{equation}
2\lambda E_{\lambda}^3+3\lambda^2 E_{\lambda}^2\frac{dE_{\lambda}}{d\lambda}-2LE_{\lambda}^2-4\lambda LE_{\lambda}\frac{dE_{\lambda}}{d\lambda}+L^2\frac{dE_{\lambda}}{d\lambda}=0.
\end{equation}
The integration with respect to $\lambda$ results in 
\begin{equation}
\lambda^2 E_{\lambda}^3 -2\lambda LE_{\lambda}^2+L^2E_{\lambda}-\alpha=0 
\end{equation}
where $\alpha$ is a constant of integration that is equal to $L^2E_0$. Therefore, we come to the expression
\begin{equation}
 E_{\lambda}=\frac{L^2E_0}{(L-\lambda  E_{\lambda})^2}. 
\end{equation}
This is exactly the spectrum of free fermions and the CSM deformed by the $T\bar{T}$ operator (see section 2).
The same formula holds for the Hamiltonian. We can rewrite it in the form
\begin{equation}\label{53}
 \textstyle\frac{1}{L}\displaystyle H_{\lambda}=f_{\lambda}(\textstyle\frac{1}{L}\displaystyle H_0),~~f_{\lambda}(x)=\frac{x}{(1-\lambda f_{\lambda}(x))^2}. 
\end{equation}


At the end of the section, let us derive the deformed classical Lagrangian of a particle moving on a certain manifold $M$ with metric $g^{\mu\nu}$. In the canonical formulation the initial Lagrangian takes the form
\begin{equation}
\mathcal{L}= p_{\mu}\dot{q}^{\mu}-\frac{g^{\mu\nu}p_{\mu}p_{\nu}}{2}
\end{equation}
where the second term is the Hamiltonian.
The deformation acts as follows
\begin{equation}
\mathcal{L}_{\lambda}= p_{\mu}\dot{q}^{\mu}-f_{\lambda}\big(\textstyle\frac{p^2}{2}\big).
\end{equation}
It is easy to find the inverse function $f_{\lambda}^{-1}(x)=x(1-\lambda x)^2$.
Thus, in the new coordinates, such that
\begin{equation}
p^{\mu}=\tilde{p}^{\mu}(1-\lambda\textstyle\frac{\tilde{p}^2}{2}),
\end{equation}
we get
\begin{equation}\label{3.33}
\mathcal{L}_{\lambda}=\tilde{p}_{\mu}\big(1-\lambda\textstyle\frac{\tilde{p}^2}{2}\big)\dot{q}^{\mu}-\textstyle\frac{\tilde{p}^2}{2},
\end{equation}
or
\begin{equation}
\mathcal{L}_{\lambda}=\tilde{p}_{\mu}\dot{q}^{\mu}-\frac{\tilde{g}^{\mu\nu}\tilde{p}_{\mu}\tilde{p}_{\nu}}{2},\,\,\,\tilde{g}^{\mu\nu}={g}^{\mu\nu}(1+\lambda \tilde{p}_{\alpha}\dot{q}^{\alpha}).
\end{equation}
The Hamiltonian equation takes the form
\begin{equation}
\dot{q}^{\mu}=\frac{\partial H_{\lambda}}{\partial p_{\mu}}=\frac{\tilde{p}^{\mu}}{1-3\lambda \textstyle\textstyle\frac{\tilde{p}^2}{2}}.
\end{equation}
Expressing $\tilde{p}$ through $\dot{q}$ and inserting it in \eqref{3.33}, we obtain the deformed Lagrangian
\begin{widetext}
\begin{equation}
\mathcal{L}_{\lambda}(\dot{{q}})=\frac{1-
\sqrt{1-6\lambda\dot{q}^2}}{3\lambda}-\frac{1}{2\dot{q}^2}\Bigg(\frac{1-
\sqrt{1-6\lambda\dot{q}^2}}{3\lambda}\Bigg)^2
+\frac{\lambda}{2\dot{q}^2}\Bigg(\frac{1-
\sqrt{1-6\lambda\dot{q}^2}}{3\lambda}\Bigg)^3.
\end{equation}
\end{widetext}

\section{\texorpdfstring{${{\mathcal{O}}}^{\text{\tiny{GHR}}}_{\lambda}$}{}-Yang-Mills theory}
As we pointed out in Introduction, the quantum two-dimensional Yang-Mills is equivalent to a system of free non-relativistic fermions, but the $T\bar{T}$  deformation spoils this relation. Therefore, it is natural to ask which deformation of the gauge theory is aligned with the $T\bar T$-perturbed system of particles. We suggest the following composite operator
\begin{equation}\label{60}
\mathcal{O}_{\lambda}^{\text{\tiny{GHR}}}(x,t)=\frac{2({T_{\lambda}}^t_{\,t}(x,t))^2+{T_{\lambda}}^x_{\,t}{T_{\lambda}}^t_{\,x}(x,t)}{1-3\lambda {T_{\lambda}}^t_{\,t}(x,t)}
\end{equation}
and use it as a definition of the new deformation of the 2d YM.

\subsection{Classical YM}
The Lagrangian density
\begin{equation}
\mathcal{L}^{\text{\tiny{YM}}}=-\frac{1}{2g^2}\text{Tr}F^2=-\frac{1}{4g^2}F_a^{\mu\nu}F^a_{\mu\nu}.
\end{equation}
For the compactness of the formulas in this section, we will omit the notation Tr of the trace over the gauge indexes.
We define the stress energy tensor as follows
\begin{equation}\label{4.3}
(T^{\text{\tiny{YM}}})^{\mu}_{\,\nu}=\frac{\partial \mathcal{L}^{\text{\tiny{YM}}}}{\partial(\partial_{\mu} \mathcal{A}^a_{\rho})}F^a_{\nu\rho}-\eta^{\mu}_{\,\nu}\mathcal{L}^{\text{\tiny{YM}}}.
\end{equation}
Therefore, the ${{\mathcal{O}}}^{\text{\tiny{GHR}}}$-YM flow equation is
\begin{equation}
\frac{d\mathcal{L}^{\text{\tiny{YM}}}_{\scriptstyle\lambda\displaystyle}(\mathbf{x})}{d\lambda}=-\frac{2\big((T^{\text{\tiny{YM}}}_{\lambda})^{t}_{\,t}\big)^2}{1-3\lambda (T^{\text{\tiny{YM}}}_{\lambda})^{t}_{\,t}}.
\end{equation}

In general, to obtain the solution $\mathcal{L}^{\text{\tiny{YM}}}_{\scriptstyle\lambda\displaystyle}$, one is instructed to substitute $(T^{\text{\tiny{YM}}})^{t}_{\,t}$ from \eqref{4.3}
and proceed to solve the differential equation perturbatively in $\lambda$ using the ansatz $\mathcal{L}_{\lambda}=\sum_{j=0}^{\infty}\lambda^jL_j$ with the initial condition $L_{0}=\mathcal{L}^{\text{\tiny{YM}}}$. However, we suggest deriving the deformed Lagrangian density by another means.

First, considering that $(T^{\text{\tiny{YM}}})^{t}_{\,t}=\mathcal{H}^{\text{\tiny{YM}}}$, we arrive at the flow equation for the Hamiltonian density
\begin{equation}
\frac{d\mathcal{H}^{\text{\tiny{YM}}}_{\scriptstyle\lambda\displaystyle}(\mathbf{x})}{d\lambda}=\frac{2\big(\mathcal{H}^{\text{\tiny{YM}}}_{\lambda}(\mathbf{x})\big)^2}{1-3\lambda \mathcal{H}^{\text{\tiny{YM}}}_{\lambda}
(\mathbf{x})},
\end{equation}
the solution of which is known from the previous section:
\begin{equation}
\mathcal{H}^{\text{\tiny{YM}}}_{\lambda}=\frac{\mathcal{H}^{\text{\tiny{YM}}}_{0}}{(1-\lambda \mathcal{H}^{\text{\tiny{YM}}}_{\lambda})^2}.
\end{equation}
After that, we rewrite the Lagrangian density in the first order formulation 
\begin{equation}
\mathcal{L}^{\text{\tiny{YM}}}=\phi F+\textstyle\frac{g^2}{2}\displaystyle\phi^2\omega
\end{equation}
where the algebra valued scalar field $\phi$ is introduced, and $\omega$ is the unit volume form. The second term here is the Hamiltonian density. In this way, the ${{\mathcal{O}}}^{\text{\tiny{GHR}}}_{\lambda}$ deformation reads as follows
\begin{equation}
\mathcal{L}^{\text{\tiny{YM}}}_{\lambda}=\phi F+f_{\lambda}\big(\textstyle\frac{g^2}{2}\displaystyle\phi^2\big).
\end{equation}
Since the scalar $\phi$ is just a Lagrangian multiplier, we are free to rename it appropriately. Thus, after the substitution
\begin{equation}
\phi\rightarrow \tilde{\phi}(1-\lambda\textstyle\frac{g^2}{2}\displaystyle\tilde{\phi}^2 ),
\end{equation}
the deformed Lagrangian density dramatically simplifies:
\begin{equation}
\mathcal{L}^{\text{\tiny{YM}}}_{\lambda}= \tilde{\phi}(1-\lambda\textstyle\frac{g^2}{2}\displaystyle\tilde{\phi}^2) F+\textstyle\frac{g^2}{2}\displaystyle\tilde{\phi}^2.
\end{equation}
Then we integrate the auxiliary field $\tilde{\phi}$, which means that we substitute a value that minimizes the action
\begin{equation}
\tilde{\phi}^*=\frac{1-\sqrt{1+6\lambda F^2/g^2}}{3\lambda F},
\end{equation}
and obtain the ${{\mathcal{O}}}^{\text{\tiny{GHR}}}_{\lambda}$-YM Lagrangian density

\begin{widetext}
\begin{equation}
\mathcal{L}^{\text{\tiny{YM}}}_{\lambda}=\frac{1-
\sqrt{1+6\lambda F^2/g^2}}{3\lambda}+\frac{g^2}{2F^2}\Bigg(\frac{1-
\sqrt{1+6\lambda F^2/g^2}}{3\lambda}\Bigg)^2\\
-\frac{\lambda g^2}{2F^2}\Bigg(\frac{1-
\sqrt{1+6\lambda F^2/g^2}}{3\lambda}\Bigg)^3.
\end{equation}
\end{widetext}
An essential characteristic of this theory is the presence of the critical value of the field strength
\begin{equation}
F^2_{\text{cr}}=-\frac{g^2}{6\lambda}.
\end{equation}

\subsection{Quantum YM}
Let us consider the YM on a cylinder with $L$-periodic space and length $T$ and specify the holonomy matrices at the endpoints of the time interval
\begin{equation}
    U(0)=e^{i\boldsymbol{\theta}_0},~~U(T)=e^{i\boldsymbol{\theta}_T}.
\end{equation}
The deformed YM's partition function can be written down as follows
\begin{multline}
Z_{\lambda}(T,e^{i\boldsymbol{\theta}_0},e^{i\boldsymbol{\theta}_T})=\int[\mathcal{D}{\mathcal{A}}(t)]\exp\Big\{-\text{Tr}\int_0^Tdt \mathcal{L}^{\text{\tiny{YM}}}_{\lambda}\Big\}\\
\times\delta\Big(\big[{P}e^{i\int_0^Tdt{\mathcal{A}}(t)}\big]e^{i\boldsymbol{\theta}_0},e^{i\boldsymbol{\theta}_T}\Big)
.
\end{multline}
Taking into account the equivalence of the Lagrangian and Hamiltonian approaches\footnote{Strictly speaking, a difference may appear because of operator-ordering related counterterms.}, one can also represent it as the quantum mechanical propagator 
 \begin{equation}
\bra{e^{i\boldsymbol{\theta}_0}}e^{-T\hat{H}^{\text{\tiny{YM}}}_{\lambda}}\ket{e^{i\boldsymbol{\theta}_T}}.
\end{equation}
The deformed Hamiltonian $\hat{H}^{\text{\tiny{YM}}}_{\lambda}$ is diagonalized in the representation basis $\ket{R}$ with wavefuctions provided by irreducible characters $\chi_{R}(e^{i\boldsymbol{\theta}})=\text{Tr}_R(e^{i\boldsymbol{\theta}})$ and the eigenvalues 
\begin{multline}
E_{\lambda}(R)=Lf_{\lambda}\Big(\textstyle\frac{g^2}{2}\displaystyle C_2(R)\Big),\\C_2(R)=\sum_{j=1}^Nn_j\big(n_j+(N+1-2j)\big)
\end{multline}
where the representations $R$ are labeled by sets of integers $\{n_1\geq n_2...\geq n_N\}$ (see \cite{Cohomology}).
Expanding the boundary states in this basis
$\ket{e^{i\boldsymbol{\theta}}}=\sum_{R}\chi_{R}(e^{i\boldsymbol{\theta}})\ket{R}$, we obtain the YM kernel
 \begin{equation}
Z_{\lambda}(T,e^{i\boldsymbol{\theta}_0},e^{i\boldsymbol{\theta}_T})=\sum_{R}\chi_{R}(e^{i\boldsymbol{\theta}_0})\chi^*_{R}(e^{i\boldsymbol{\theta}_T})e^{-Af_{\lambda}(C_2(R))}.
\end{equation}
The area of the cylinder $A=LT$.
Note that the Hilbert space has not changed, and no extra eigenvalues have appeared. 
Rewriting the characters in the following way
\begin{multline}
\chi_R(e^{i\boldsymbol{\theta}})=\det\limits_{_{ab}}
 \big(e^{i (n_a-a)\theta_b}\big)
/{\Delta}(e^{i\boldsymbol{\theta}}),\\{\Delta}(e^{i\boldsymbol{\theta}})=\prod\limits_{s<r}(e^{i\theta_s}-e^{i\theta_r})
\end{multline}
we find that the deformed partition function satisfies the differential equation
\begin{multline}
\Big[\big(\textstyle\frac{2}{N^2}\displaystyle\partial_A-\textstyle\frac{1}{12}\displaystyle\big)\big(1-\textstyle\frac{\lambda}{2}\displaystyle\big(\textstyle\frac{2}{N^2}\displaystyle\partial_A-\textstyle\frac{1}{12}\displaystyle\big)\big)^2\\
-\frac{1}{{\Delta}(e^{i\boldsymbol{\theta}})}\sum_{i=1}^N\frac{\partial^2}{\partial \theta^2_i}{\Delta}(e^{i\boldsymbol{\theta}})\Big]Z_{\lambda}=0
\end{multline}
where the first term is $f_{\lambda}^{-1}\big(\frac{2}{N^2}\partial_A-\frac{1}{12}\big)$.

\section{Hamiltonian reduction: deformed YM+Wilson line}

The pure YM on the cylinder is equivalent to the unitary matrix quantum mechanics describing free fermions on a circle. The addition of a massive color source along the time direction results in placing the fermions in the trigonometric Calogero-Sutherland potential.
In the present section, we investigate how this relation behaves under the $T\bar{T}$ and $\mathcal{O}^{\text{\tiny{GHR}}}$ deformations. We show that these deformations are completely different before the Hamiltonian reduction, but they coincide on the reduced phase space.

    Consider $U(N)$ YM on the cylinder with periodic space coordinate.  The action in the canonical formulation is
\begin{equation}
S^{\text{\tiny{YM}}}=\int_{S^1\times \mathbb{R}}\text{Tr}\big(\phi F+\textstyle\frac{g^2}{2}\displaystyle\phi^2\big).
\end{equation}
The scalar field $\phi$ represents an electric field. Note that the solvability of 2d YM lies in the absence of a magnetic field in 1+1 dimensions.
We can rewrite the Lagrangian as follows
\begin{equation}
\mathcal{L}^{\text{\tiny{YM}}}=\text{Tr}\big(2\phi\partial_t A_x+\textstyle\frac{g^2}{2}\displaystyle\phi^2+A_tD_x\phi\big).
\end{equation}
The time component $A_t$ in the last term plays the role of a Lagrange multiplier for the Gauss law and the momentum map constraint
\begin{equation}
\mu=D_x\phi=\partial_x\phi+[A_x,\phi
]=0.
\end{equation}
This ensures that the electric field is covariantly constant.

\subsection{Deforming before the reduction}
The first and the third terms in the Lagrangian are topological, while the second one involves the volume form and transforms under these two deformations in the following ways
\begin{equation}\nonumber
T\bar{T}:~~g^2\text{Tr}\phi^2\rightarrow\frac{\frac{g^2}{2}\text{Tr}\phi^2}{1-\lambda \frac{g^2}{2}\text{Tr} \phi^2},
\end{equation}
\begin{equation}
\mathcal{O}^{\text{\tiny{GHR}}}:~~g^2\text{Tr}\phi^2\rightarrow f_{\lambda}(g^2\text{Tr}\phi^2),
\end{equation}
where the function $f_{\lambda}(x)$ is defined in \eqref{53}.

After the deformation is performed, we insert the Wilson line in certain representation $R_{\gamma}$ 
along the time direction
\begin{equation}
\bra{R_{\gamma}^*}\text{Tr}_{R_{\gamma}}\big(P\exp{\int_t A_t}\big)\ket{R_{\gamma}}.
\end{equation}
If the time-like Wilson lines, i.e., electrical sources, are inserted at points $x_i$ of the spatial circle, the Gauss law changes in the following way
\begin{equation}\label{85}
\mu+\sum_i\mu_{\mathcal{O}_i}\delta(x-x_i)=0,
\end{equation}
where $\mu_{\mathcal{O}_i}$ are momentum maps corresponding to the finite dimensional orbits ${\mathcal{O}_i}$ located at the points $x_i$. 
To get spinless Calogero-Sutherland dynamics on the reduced phase space, we leave only one Wilson line at $x_1$ in the one-row representation $R_{\gamma}$ \footnote{The corresponding Young tableau is a row of $N\gamma$ boxes.}. The respective orbit is the complex projective space $CP^{N-1}$. In the homogenious coordinates $z,\,\bar{z}$ the momentum map is as follows $\mu_{\scriptscriptstyle CP^{{N-1}}}=\gamma(z\otimes\bar{z}-\mathbb{1})=J_{\gamma}$. Therefore we obtain
\begin{multline}
S_{\substack{\scriptscriptstyle T\bar{T}\text{-YM}\\ \text{\tiny{+WL}}}}=\int_{S^1\times T}dx dt\,\text{Tr}\Big[ \phi\partial_{t}A_{x} +\frac{\frac{g^2}{2}\phi^2}{1-\lambda \frac{g^2}{2}\text{Tr} \phi^2}\\
-A_t\big(\partial_{x}\phi+[A_x,\phi]-\delta(x-x_1)J_{\gamma}\big)\Big].
\end{multline}
The similar expression holds for the $\mathcal{O}^{\text{\tiny{GHR}}}$. 

Consider the gauge $A_x=\text{diag}\{i\theta_1,...,i\theta_N\}$. Therefore, the solution of the Gauss law \eqref{85} outside the point $x_1$ is as follows
\begin{equation}\label{87}
\phi_{ij}(x)=e^{- i(x-x_1)(\theta_i-\theta_j)}\phi_{ij}(x_1).
\end{equation}
Moreover, the field jumps at point $x_1$
\begin{equation}\label{88}
\phi_{ij}(x_1+0)-\phi_{ij}(x_1-0)=\mu_{\scriptscriptstyle CP^{{N-1}}}.
\end{equation}
Using \eqref{87} and \eqref{88} we come to the conditions
\begin{multline}
(\mu_{\scriptscriptstyle CP^{{N-1}}})_{ii}=0~\rightarrow~|z|^2=1,\\
\phi_{ij}(x_1+0)=\frac{\gamma z_i\bar{z}_j}{1-e^{-2\pi i(\theta_i-\theta_j)}},\\
\phi_{ii}= p_i=const.
\end{multline}
Ultimately, on a subspace satisfying these conditions, we obtain the Hamiltonian of the reduced theory:
\begin{equation}
H_{\lambda}^{\text{\tiny{GHR}}}=\int dx f_{\lambda}\big(\textstyle\frac{g^2}{2}\displaystyle\text{Tr}\phi^2\big)=Lf_{\lambda}\big(\textstyle\frac{1}{L}\displaystyle H^{\text{\tiny{CSM}}}\big),
\end{equation}
\begin{equation}
H_{\lambda}^{T\bar T}=\frac{ H^{\text{\tiny{CSM}}}}{1-\lambda H^{\text{\tiny{CSM}}}/L},
\end{equation}
where
\begin{equation}
H^{\text{\tiny{CSM}}}=\frac{g^2L}{2}\bigg(\sum_{i=1}^Np_i^2+\sum_{i<j}^N\frac{\gamma^2}{\sin^2\frac{\theta_i-\theta_j}{2}}\bigg)
\end{equation}
is the Hamiltonian of Calogero-Sutherland particles living on a circle of length ${L^{\prime}}=\frac{\pi}{\sqrt{g^2L}}$. 

So far, we have described the Hamiltonian reduction of deformed theories with an additional Wilson line. Also, one can add the Wilson line before performing the deformations. We will elaborate on this case elsewhere. This reduction can also be repeated in the quantum case. As a result, the quantum shift appears $\gamma^2\rightarrow\hbar^2\beta(\beta-1)$ (see \cite{gn2}).


\subsection{Deforming after the reduction}
Recall that by imposing the Gauss law constraint on the pure Yang-Mills theory, one obtains the dynamics of free fermionic particles, while the YM+WL system gets reduced to the CSM.
According to the definition of the $\mathcal{O}^{\text{\tiny{GHR}}}$ deformation, the Hamiltonians change as follows
\begin{equation}
\textstyle\frac{1}{L}\displaystyle H^{\text{free}}\rightarrow f_{\lambda}\big(\textstyle\frac{1}{L}\displaystyle H^{\text{free}}\big),~~
\textstyle\frac{1}{L}\displaystyle H^{\text{\tiny{CSM}}}\rightarrow f_{\lambda}\big(\textstyle\frac{1}{L}\displaystyle H^{\text{\tiny{CSM}}}\big).
\end{equation}

\begin{figure*}[t]
\includegraphics[scale=0.43]{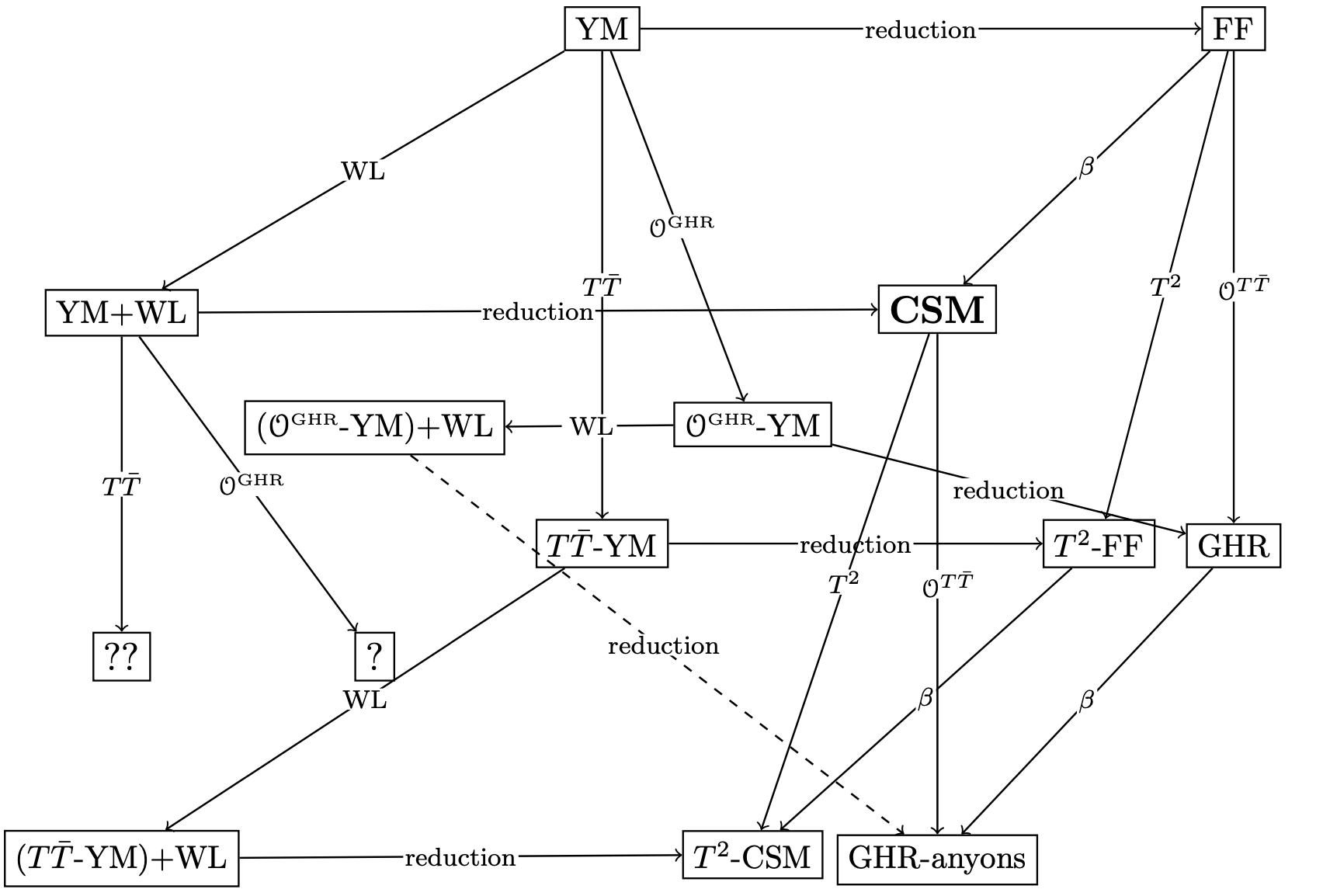}
\caption{\label{fig:1}The road map of the deformations. The arrow ``WL'' stands for insertion of the Wilson line along the cylinder in the one-row representation; ``reduction''$-$ the Hamiltonian reduction; the dashed arrow indicates that the correspondence takes place only between the spectra; ``GHR''$-$generalized hard-rod system; ``GHR-anyons''$-$ GHR with anyonic statistics; ``FF'' $-$free fermions; $T^2$$-$deformation of the QM by the square of ${T}^t_{~t}$; $\mathcal{O}^{T\bar{T}}$$-$bilinear operator \eqref{2.22}; $\mathcal{O}^{\text{\tiny{GHR}}}$ is defined in \eqref{60}; $\beta-$the quantum coupling constant of CSM. The CSM degenerates to the FF at $\beta=1$.}
\label{ris:image3}
\end{figure*}

In the quantum case, as we described in section 2, the $T\bar{T}$ deformation leads to the
same Hamiltonians up to the canonical transformation
\begin{equation}
\textstyle\frac{1}{L}\displaystyle H^{\text{free}}\rightarrow f_{\lambda}\big(\textstyle\frac{1}{L}\displaystyle H^{\text{free}}\big),
\end{equation}
\begin{equation}
\textstyle\frac{1}{L}\displaystyle H^{\text{\tiny{CSM}}}\rightarrow U_{\lambda}f_{\lambda}\big(\textstyle\frac{1}{L}\displaystyle H^{\text{\tiny{CSM}}}\big)U^{-1}_{\lambda}.
\end{equation}
From the point of view of one-dimensional theory, we consider a classical particle on the manifold $M$. When the Wilson line is added, the phase space expands
\begin{equation}
T^*M\rightarrow T^*M\times CP^{N-1}.
\end{equation}
The particle acquires additional degrees of freedom respective to the motion in $CP^{N-1}$. So the Lagrangian of the particle becomes
\begin{equation}
\mathcal{L}^{\text{free\tiny{+WL}}}= p_{\mu}\dot{q}^{\mu}- H^{\text{free}}+\theta-\text{Tr}(A_t\,\mu
_{\scriptscriptstyle CP^{{N-1}}})
\end{equation}
where $d\theta=\gamma\,\omega_{FS}$ -- the Fubini-Study form, and $\int \theta$ is the Berry phase.
Deforming this system we obtain 
\begin{equation}
H_{\lambda}^{\text{free\tiny{+WL}}}=f_{\lambda}\big(H^{\text{free}}-\theta\,+\text{Tr}(A_t\,\mu_{\scriptscriptstyle CP^{{N-1}}})\big).
\end{equation}
We summarize the results of the section in FIG. \ref{fig:1}.

\section{Discussion}
In this work, on the example of free particles and the Calogero-Sutherland model, we proposed the new $T\bar T$-like deformation of non-relativistic many-body quantum mechanics. We derived it by dimensional reduction of the bilinear $T\bar T$ operator using the generalized trace relation. The deformation alters the spectrum of a Hamiltonian but does not change its eigenfunctions. Furthermore, the deformed Lagrangian of non-interacting particles was also obtained. Finally, we described the deformed 2d Yang-Mills theory that is equivalent to a system of $T\bar T$-fermions.


There is a range of exciting directions for further study. One of them concerns the possible geometrisation of $\mathcal{O}^{\text{\tiny{GHR}}}$ deformation. As is well known, there is an interesting geometric interpretation of a $ T \bar T $-perturbed two-dimensional QFT. So the deformation can be described as coupling the theory to the Jackiw-Teitelboim gravity \cite{DubGorbMirb,geom1}. Recently, the same interpretation was found for the case of non-Lorentz invariant QFTs \cite{geom2}. In \cite{gross1,gross2}, the deformation of quantum mechanical theories was also described as coupling a seed QM to a specific worldline gravity. It would be instructive to formulate our deformation in a similar way.

Moreover, given the natural appearance of the $\mathcal{O}^{\text{\tiny{GHR}}}$ operator,
it seems worthwhile to further explore this deformation in itself. Of particular interest is the deformation of the Schwarzian quantum mechanics and the subsequent clarification of its possible holographic interpretation. Although the new operator $\mathcal{O}^{\text{\tiny{GHR}}}$ and the operator proposed in \cite{gross1} look very similar, they lead to significantly different deformed theories. So
the spectral density of the $\mathcal{O}^{\text{\tiny{GHR}}}$-perturbed Schwarzian theory is
\begin{equation}
\rho_{\lambda}(E)=\frac{\big(1-4\lambda E+3\lambda^2E^2\big)}{\sqrt{2\pi^3}}\sinh{\Big(2\pi\sqrt{2E(1-\lambda E)^2}\Big)}.
\end{equation}
It is well defined over the entire range of $\lambda$ and demonstrates super-Hagedorn behavior\footnote{The problem of having a complex spectrum in the deformed Schwarzian theory was also addressed in \cite{add2}.}.

Another interesting direction is studying the $T\bar T$ deformation of a system of Brownian particles on a circle. The problem of non-intersecting Brownian walks has been highly instructive and relevant in many issues. The authors of \cite{Deh,fms1,fms2}  discovered an unexpected connection between 2d YM on a sphere and Brownian motions. They showed the mapping of the  YM partition function to the normalized reunion probabilities of free non-intersecting Brownian walks on a line with different boundary conditions.
These results were recently generalized to the case of the YM on a cylinder and random walkers interacting via the trigonometric Calogero-Sutherland potential \cite{gmn}.
 Using the hard rod intuition, we are going to study the YM/random walks correspondence in the $T\bar T$ setting for both cases of free and interacting Brownian process. 

Finally, we would like to draw attention to the close relation between the 2d YM and string theory. It was shown a long time ago that the partition function of the YM on a sphere admits a dual string interpretation \cite{GrossTaylor}. It is based on the factorization of the gauge theory into chiral and anti-chiral sectors in the large $N$ limit, which follows immediately from the fermionic picture. The $T\bar T$-YM/string theory correspondence has been recently discussed \cite{Shyam,tierz2}.  It has been shown that the  $T\bar T$-YM theory is mapped onto a system of fermions experiencing highly non-local interaction that prevents their factorization. Thus, it casts doubt on a possible string-like formulation. On the other hand, 
the $\mathcal{O}^{\text{\tiny{GHR}}}$-perturbed YM discussed in the current paper has the clear interpretation in terms of free generalized hard-rod fermions and therefore deserves further investigation in this direction.

\begin{acknowledgments}
I am grateful to A.S. Gorsky for inspiring discussions and collaborations on related projects. The work was supported by BASIS Foundation grant 20-1-1-23-1 and the grant RFBR-19-02-00214.
\end{acknowledgments}

\end{document}